\documentclass[12pt,preprint]{aastex}
\shorttitle{An Infrared-Red Nucleus in NGC 315} \shortauthors{Gu,
Huang, Wilson, \& Fazio}

\def\kms{\ifmmode {\rm km~ s^{-1}} \else {\rm km~s^{-1}}\ \fi}

\begin{document}

\title{Direct Evidence from {\it Spitzer} for a low-luminosity AGN at the center
of the Elliptical Galaxy NGC 315}

\author{Q.-S. Gu\altaffilmark{1,2},
J.-S. Huang\altaffilmark{2}, G. Wilson\altaffilmark{3} and G. G.
Fazio\altaffilmark{2}}

\altaffiltext{1}{Department of Astronomy, Nanjing University,
Nanjing 210093, China; qsgu@nju.edu.cn}

\altaffiltext{2}{Harvard-Smithsonian Center for Astrophysics, 60
Garden Street, Cambridge, MA 02138}

\altaffiltext{3}{Spitzer Science Center, California Institute of
Technology, MS 220-6, Pasadena, CA 91125}

\begin{abstract}
We present the {\it Spitzer}  Space Telescope InfraRed Array Camera (IRAC) 
and Multiband Imaging Photometer (MIPS) observations of the elliptical
galaxy NGC~315.  After removal of the host galaxy's stellar 
 emission, we detected for the first time an infrared-red nucleus in NGC~315. 
 We measured the spectral energy distribution (SED) for this active nucleus 
 with wavelength range covering from radio to X-ray, and obtained the 
 bolometric luminosity of $\rm L_{bol} \approx 1.9 \times 10^{43} ergs \
 s^{-1}$, corresponding to an extremely low Eddington ratio (L/L$_{\rm Edd}$) of 4.97 $\times$
 10$^{-4}$. Our results confirm that the physical nature of the
 nucleus of NGC 315 is a low-luminosity AGN, consistent with the
 recent optical and {\it Chandra} X-ray observations.

\end{abstract}


\section{Introduction}

 The recent discovery of a tight correlation between black hole (BH)
 mass and the bulge's mass of host galaxy suggests that the creation
 and growth of central black holes are an integral part of the formation
 of galactic bulges (Kormendy \& Richstone 1995; Kauffmann \& Heckman 2005). On the other hand,
 it also implies that supermassive black
 holes (SMBHs) are ubiquitous at the centers of elliptical galaxies. Although indeed most
 host galaxies of radio-loud and the brightest
 quasars at high-redshift are elliptical galaxies (Hutchings \& Morris 1995;
 Bahcall, Kirkakos \& Schneider 1996; Falomo et al. 2005), the
 majority of SMBHs in nearby ellipticals are not very active
 (Ho, Filippenko \& Sargent 1997), and  therefore hardly detectable.

 The low state of SMBHs in nearby ellipticals is most probably due
 to the lack of enough fuel supply - cold gas inflow into the central engine. 
 It is generally accepted that elliptical galaxies are spheroidal distributions of older stars
 %without much of the
 and lack substantial cold intersteller material (ISM).  
 However, an H$_\alpha +$ [NII] narrow-band imaging survey
 has shown that elliptical galaxies certainly have plenty of extended 
 warm ($\sim 10^4$ K) gas (Shields 1991). By detecting positive signal from co-adding
 images of the {\it Infrared Astronomy Satellite} (IRAS) survey scans, Knapp
 et al. (1989) discovered that $\sim$ 50 percent of optically-selected
 ellipticals with visual magnitude brighter than V=14 contained
 cool interstellar matter.  
 Temi et al. (2004) estimated that 16
 elliptical galaxies observed with the {\it Infrared Space
 Observatory} (ISO) have dust masses on average 10 times greater
 than those previously detected by IRAS.
 Most recently, Kaneda, Onaka, \& Sakon (2005)  detected even polycyclic
 aromatic hydrocarbon (PAH) emission features in four elliptical
 galaxies with the {\it Spitzer} Infrared Spectrograph (IRS). 
 Recent neutral hydrogen observations also show that
 plenty of extended gas is clearly present in early-type galaxies (Morganti et al. 2006; 
 Noordermeer 2006 and references therein). 
 All these results suggest that the presence of a significant amount of cold/warm
 interstellar gas at centers of elliptical galaxies is quite common,
 although the origin of such amount of dust and cold gas is still uncertain
 (Caon et al. 2000).

 AGN signature of SMBHs has been seen in elliptical galaxies.
 In the recent Palomar spectroscopic survey, Ho, Filippenko \& Sargent (1997)
 found that about 50\% elliptical galaxies show detectable emission-line
 nuclei,  most of which ($\sim$ 87\%) are classified as
 LINERs. Although it is still a subject of hot debate whether LINERs are
 genuine low-luminosity AGNs (see the recent review by Ho 2003),
 AGN activity is confirmed by radio and X-ray observations
 (Terashima et al. 2002; Filho et al. 2004).

 Typical AGNs are also strong infrared sources. Ho et al. (1997) found that about 55\% of elliptical
 galaxies with emission lines were detected to have infrared emission by IRAS. Due to the lower
 spatial resolution ($\sim$ 3'$-$4' ) of IRAS, it is impossible to determine
whether  the far-infrared emission from elliptical galaxies arises from AGN point
 sources at the center, or from extended dust distribution seen as dust-lane in optical images (Lauer et al. 2005).
 One example is NGC 315, a well studied elliptical galaxy classified as a LINERs
 by Ho et al. (1997). The HST image
 shows a clear dusty disk with 2.5" diameter in its center (Verdoes Kleijn et al. 1999) and the galaxy is also detected
 by IRAS at 12, 25, 60, and 100 \micron\ with flux densities of 0.081, 0.150, 0.270, and 0.706 Jy,
 respectively (IRAS Faint Source Catalogue, Moshir et al. 1990), corresponding to $L_{\rm FIR}=2.3 \times 10^9 L_\odot$
 at the distance of 65.8 Mpc (Ho, Filippenko \& Sargent 1997). Clearly, a high-resolution infrared
 image of NGC 315 is required to determine whether its infrared emission comes from the central 2" region.

 The {\it Spitzer Space Telescope} (Werner et al. 2004) has 2" spatial resolution in mid-infrared
 (3.6\micron\ $-$ 8.0 \micron) and 5" at 24\micron. This makes it possible to study the infrared
 emission from central regions of  elliptical galaxies.
 In this work we present a study of the nuclear activity in the elliptical
 galaxy NGC 315 based on observations by the Infrared Array Camera (IRAC, Fazio et al. 2004) and
 the Multiband Imaging Photometer for Spitzer (MIPS, Reike et al. 2004) aboard the {\it Spitzer} Space Telescope.
 This paper is organized as follows: the IRAC and MIPS data reduction, flux calibration
 and photometry are in Sect. 2.  We present multi-wavelength physical properties of
 nucleus of NGC 315  in Sect. 3.

\section{Data Reduction}

\begin{figure}
\plotone{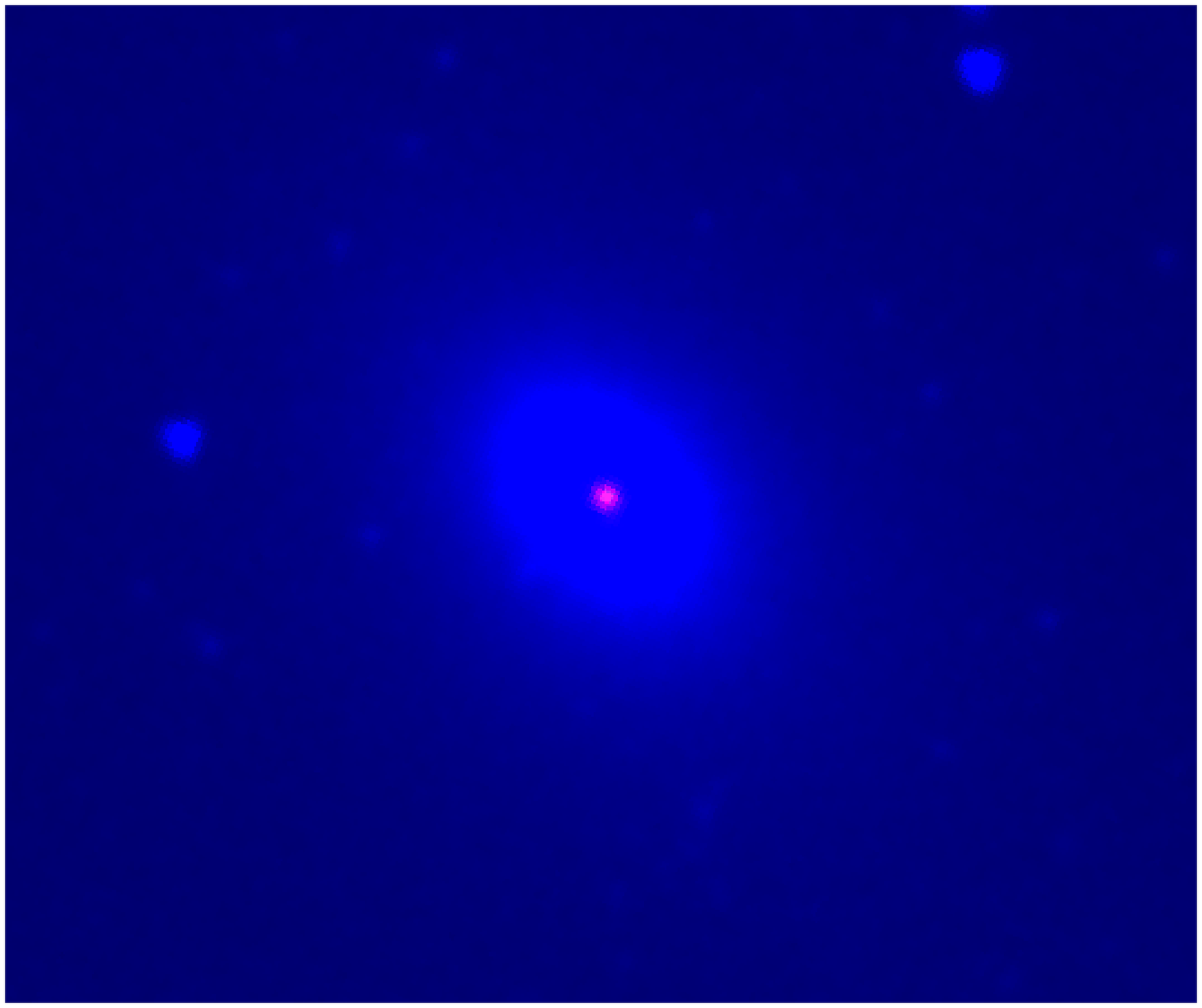} \caption{The RGB false-color image of NGC 315, which combines
IRAC 4.5$\mu$m (B), 5.8$\mu$m (G), and 8.0$\mu$m (R) images from the
{\it Spitzer Space Telescope}. A infrared-red core is clearly seen in the central
region. The image size is 2'.0 $\times$ 2'.0 \label{f:fig1}}
\end{figure}

 We downloaded the IRAC Basic Calibrated Data (BCD) for NGC 315
 from the archive of the Spitzer Science Center (SSC). The BCD images had been
 performed with basic image processing, including dark
 subtraction, detector linearization corrections, flat-field
 corrections, and flux calibrations (see also the IRAC Data
 Handbook\footnote{http://ssc.spitzer.caltech.edu/irac/dh/}). We
 further used the custom IDL software (Huang et al. 2004) to make the final
 mosaic images  as shown in Figure~\ref{f:fig1}.
 The MIPS image for NGC 315 was mosaicked by the SSC pipeline as post-BCD products.
 Both the IRAC and MIPS bands have absolute flux calibration accuracies of better than 10\% (Fazio
 et al. 2004; Rieke et al. 2004). Throughout this paper, the magnitudes
 and colors were given in the AB system.
 
 The surface photometry for NGC 315 was performed with the {\it
 ellipse} program in the ISOPHOT package of IRAF\footnote{IRAF is
 distributed by the National Optical Astronomy Observatories, which
 are operated by the Association of Universities for Research in
 Astronomy, Inc., under cooperative agreement with the National
 Science Foundation.}. The isophotal parameters, such as ellipticity
 and position angle, for NGC315 were measured at 3.6\micron\ where
 the Signal-to-Noise (S/N) ratio is the highest. These parameters were
 then applied for the surface photometry at 4.5, 5.8 and 8.0
 \micron. IRAC photometry calibration is designed only for point
 sources. Recently Tom Jarrett provided addional
 aperture corrections for extended sources photometry in IRAC
 images\footnote{http://spider.ipac.caltech.edu/staff/jarrett/irac/calibration/index.html}. 
 We applied this addiotional correction for the IRAC surface
 photometry of NGC 315. The total flux densities for NGC 315 in each
 IRAC band are given in Table~\ref{t:flux}. We diluted the 3.6, 4.5
 and 5.8\micron\ IRAC images so that they have the same spatial
 resolution as the 8.0\micron\ image to obtain accurate IRAC color
 distributions for NGC 315.

\begin{figure}
\plotone{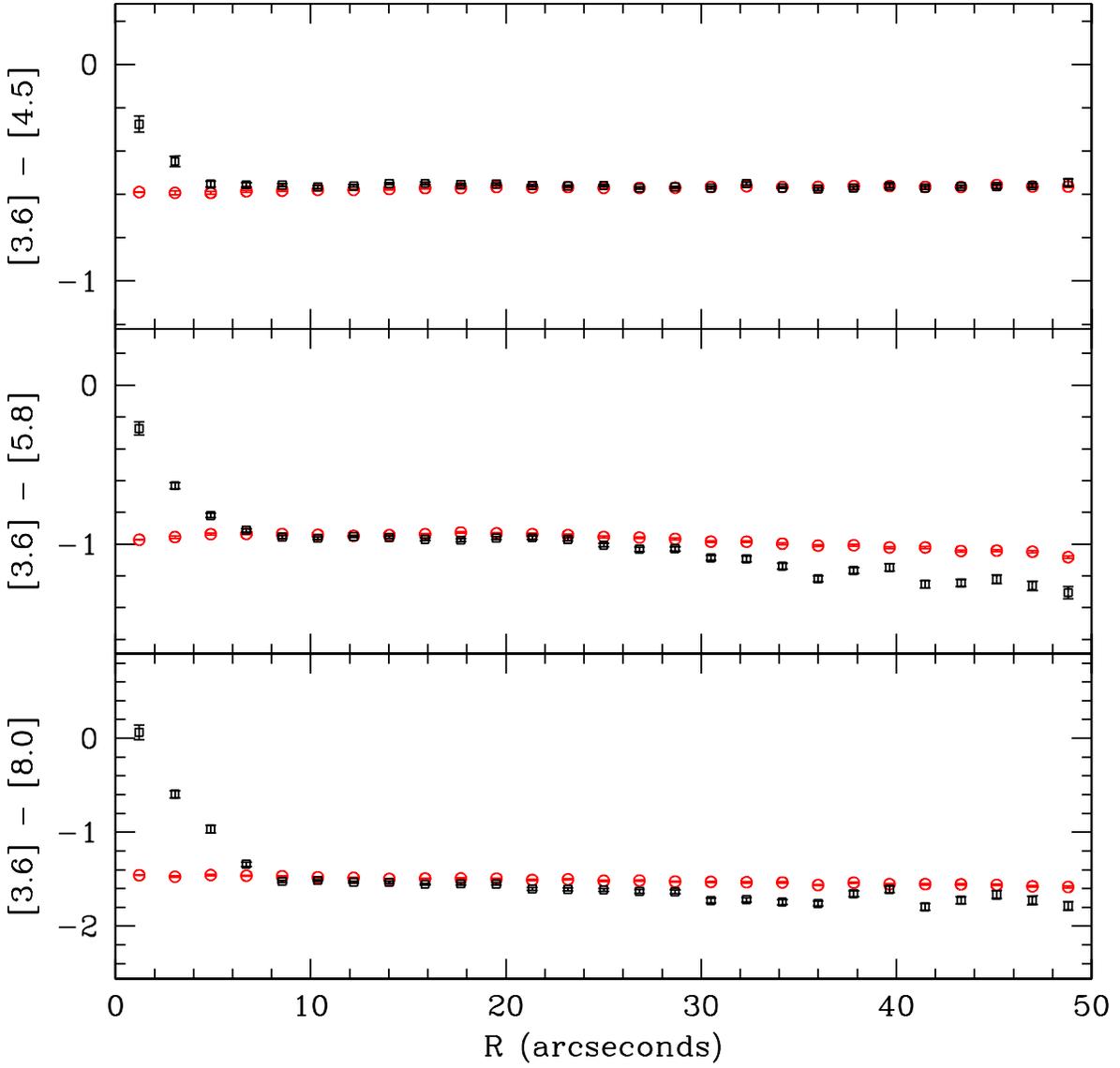} \caption{The IRAC color distribution for NGC 315. \label{f:fig2}}
\end{figure}

 Figure~\ref{f:fig2} shows three IRAC color distributions of  [3.6]-[4.5], [3.6]-[5.8], and
 [3.6]-[8.0] for NGC 315.  At 5" $<$ R $<$ 50", these infrared colors are roughly
 constant and consistent with those of an M star (Pahre, et al.
 2004). The central region, however, shows a much redder color
 indicating a non-stellar origin of the infrared emission. In
 Figure~\ref{f:fig2}, we also plot the IRAC colors of a normal
 elliptical galaxy, NGC 5557, for comparison, NGC 5557 has not been
 identified to have any AGN signature in the center, it does not
 show red IRAC colors in the central region (R $<$5").  The constant
 color distributions for NGC 315 at R $>$ 5" suggest a uniform
 stellar population, which is confirmed by the high S/N long-slit optical
 spectrophotometry (see Fig. 3 in Cid Fernandes et al. 2005).

\begin{figure}
\plotone{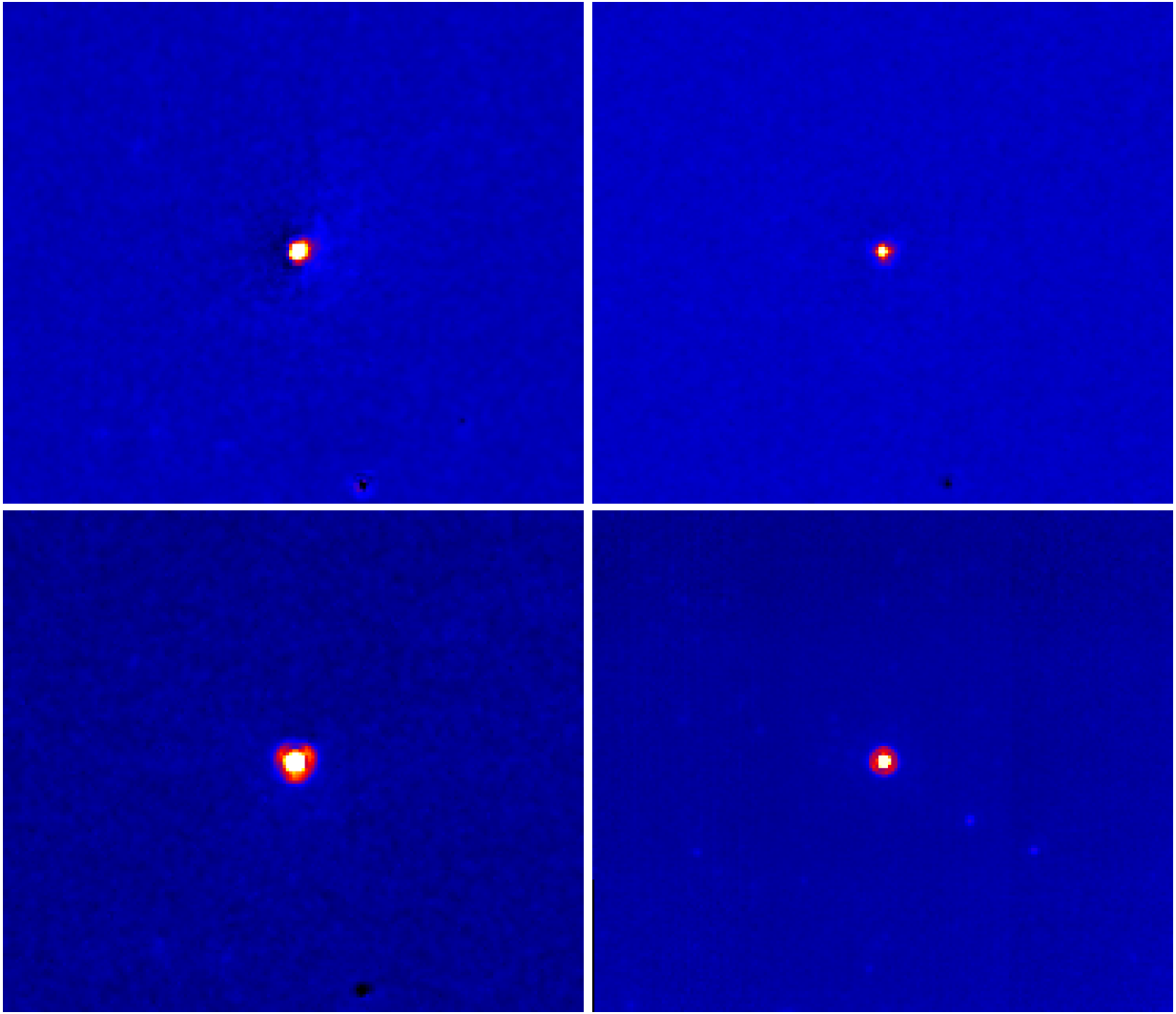} \caption{The residual images at 4.5 (upper left), 5.8 (upper right), and
 8.0 $\mu$m (lower left) of NGC 315 after
 removing the underlying stellar contribution. The 24 $\mu$m MIPS
 image is also shown in the lower-right panel.
 We can even see the ring-like structure in the residual 8.0$\mu$m image, which is
 exactly the characteristic feature of the IRAC 8.0$\mu$m Point Spread
 Function (PSF).\label{f:fig3}}
\end{figure}

 In the IRAC band the emission from the central region of NGC 315 consists of two components:
 stellar and non-stellar emission. However, we cannot interpolate the stellar distribution measured in the outer
 region to the region with R$<$2", because the stellar distribution in the central region of an elliptical
 galaxy can have various profiles. Pahre et al. (2004) argued that
 the 3.6\micron\  emission traces the stellar mass distribution very well (see also Wu et al. 2005),
 we relied on the IRAC 3.6\micron\  image to remove the contribution
 from the underlying stellar population.
We used the mean colors ([3.6]-[4.5], [3.6]-[5.8], and [3.6]-[8.0]) at
 the region of 10 and 30
 arcseconds to scale the 3.6\micron\  image to remove the contribution from the underlying stellar components
 in the 4.5, 5.8, and 8.0\micron\ images. There is only one point source left in all 3 residual images,
 as shown in Figure~\ref{f:fig3}. The flux density for the point source is as following:

 \begin{equation}
   f_{residual}(\lambda)=f_{ns}(\lambda)-R_{stellar} \times f_{ns}(3.6)
 \end{equation}

 \noindent where $\lambda$=4.5, 5.8, and 8.0\micron, f$_{residual}$ is the total flux for the point source
 in the residual images, f$_{ns}$ is the flux density for the non-stellar red nucleus, and R$_{stellar}$
 is the flux ratio of the stellar emission. We have to assume the shape of the SED for the central
 non-stellar emission to solve the 4 flux densities out of 3 equations. f$_{residual}(\lambda)$ appears
 as a power-law distribution, it is reasonable to assume that f$_{ns}(\lambda)$ is a power-law function.
 The f$_{ns}$ for 3.6, 4.5, 5.8 and 8.0\micron\ are given in Table 1. The residual point source is also
 consistent with the MIPS 24\micron\ image. NGC 315 appears as a point source at 24
 \micron\ (Figure~\ref{f:fig3}) because its stellar emission is negligible in this band.
 Therefore for the first time we determine that all infrared emission from NGC 315 detected
 by IRAS and Spitzer comes from the central point-like source.
 
 % NGC 315 was also observed in the spectral mapping mode with the Spitzer IRS using short, low resolution  
 % module (SL2). The total exposure time per slit was 14.68 sec. The pipeline BCD products from the 
 % Spitzer Science Center were detected and cleaned hot pixels with an interactive IDL tool of 
 % IRSCLEAN$\_$MASK,  and then processed within the SMART software package 
 % (Ver. 6.2.5, Higdon et al. 2004).  The nuclear spectrum was extracted using a 3.0 pixel aperture 
 % after subtracting the outmost image as background. The nuclear SL2 spectrum of NGC~315 was
 % shown in  Figure 4. The signal-to-noise (S/N) of the spectrum is rather low, the most significant feature is
 % a pronounced broad [Ar II] emission at 6.99 \micron. Such similar feature was also observed in
 % the elliptical galaxy NGC 2974 (Kaneda, Onaka \& Sakon 2005).   However, the present IRS spectrum
 % is too weak to be analyzed in detail.

\section{Physical Properties of Red Nucleus}

At the resolution of IRAC it is not possible to tell whether the IR emission is really coming from the AGN nucleus, or from the dusty disk seen in the HST images.  If the IR emission is produced by heated dust, we can estimate the dust temperature 
using the 8.0 and 24 \micron\ fluxes, since  {\it Spitzer} Infrared Spectrograph (IRS)
observations suggest that the PAH feature at 7.7 \micron\ is very faint or even 
absent in elliptical galaxies (Kaneda, Onaka \& Sakon 2005).
Assuming a grey-body radiation characterized by a dust emissivity going as  
$\lambda^{-1.5}$,  we derive a  dust temperature of  $T_{dust}\sim168$K, 
which is  similar to the dust temperatures measured in the central regions of Seyfert  galaxies, $T_{dust}\sim170$K
(Deo 2007). We also need to check whether the amount of dust mass  inferred 
from the optical dusty disk is consistent with the dust mass required to explain
the IR emission.  de Ruiter et al. (2002) estimated a dust
mass from IRAS data ($M_{dust}\sim 8.0\ 10^5 M_\odot$) that is a thousand times larger than 
that inferred from the optical reddening absorption study ($M_{dust}\sim 8.0 \ 10^2 M_\odot$),
which might indicate a significant contribution of IR emission from the central AGN. When
A ring-like structure is clearly visible in the 8.0 \micron\ image (lower-left panel of Fig. 3), which is very similar to the IRAC 8.0 Point Spread Function (PSF) first Airy ring, thus 
providing further evidence of the IR emission comes from the AGN nucleus. We conclude therefore that both the dusty disk and the central AGN contribute to the IR emission. With the current data set, we cannot address it further. For the future, 
IR variability data would certainly help to solve the current issue. 

\begin{figure}
\plotone{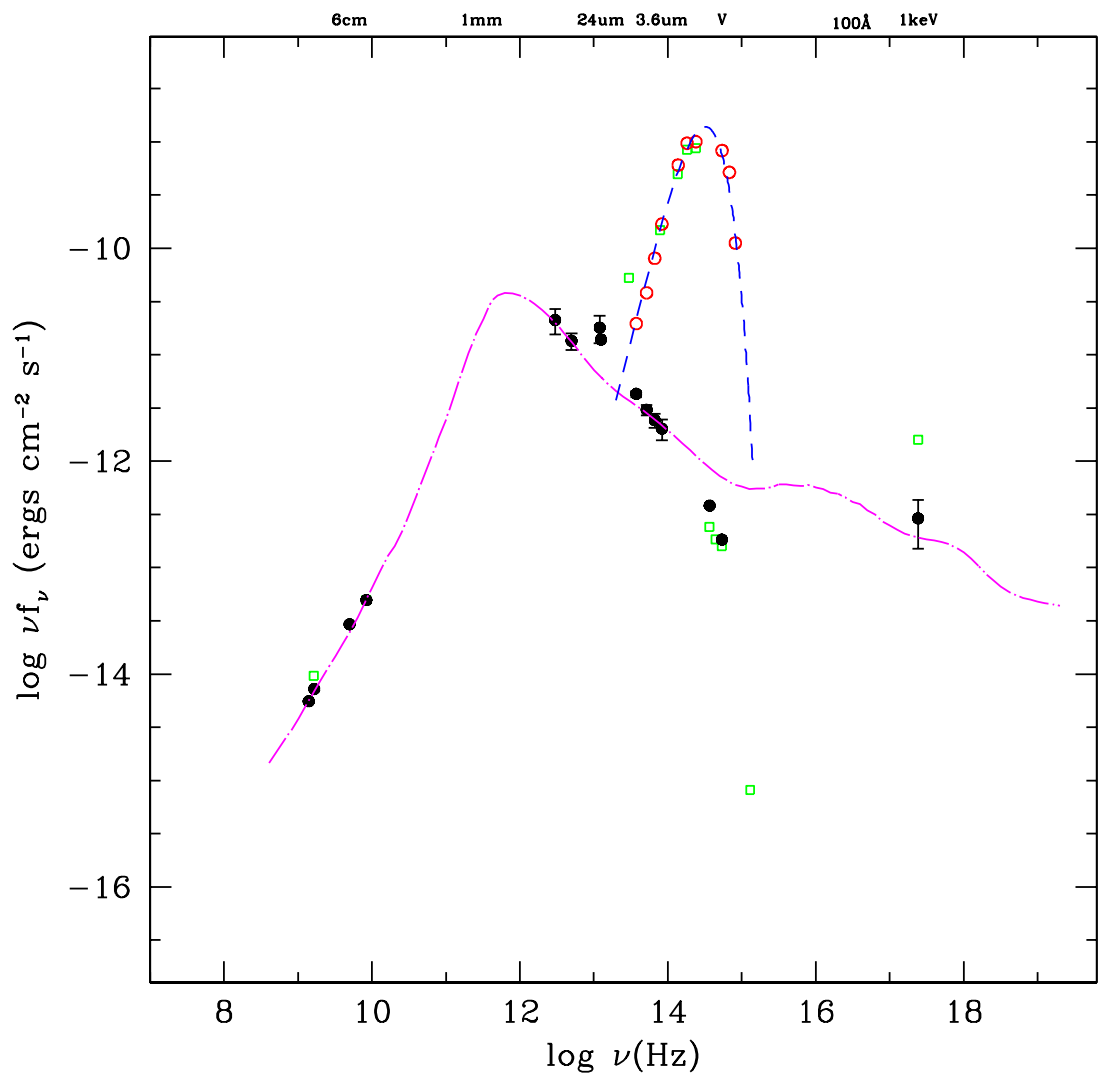} \caption{Spectral energy distribution of the infrared-red core
 in NGC~315. The SED of Sgr~A* (dot-dashed magenta line) and NGC 4261 (green squares) are
 also shown and normalized to the 8.42 GHz emission flux of NGC~315. 
 The SED of the host galaxy of NGC~315 is
 shown as red circles, which is well fitted by the black-body radiation with the temperature
 of 3950 K (dashed blue line). \label{f:fig4}}
\end{figure}

 Venturi et al. (1993) presented multifrequency radio observations of NGC 315 with parsec
 resolution, they detected radio emission from a nucleus with a size of 0.05 pc and
 $\rm \nu L_\nu(5GHz) \approx 1.6 \times 10^{40} ergs \ s^{-1}$.
 The {\it Chandra} also detected a point-like nucleus, which
 is well fitted with a single power law  with an intrinsic absorption of 
 $\rm n_{\rm H} \sim (7.6\pm 1.2) \times 10^{21} cm^{-2}$,  and an 0.5-8 keV luminosity of
 of $(7.6 \pm 0.3) \times 10^{41} \ erg \ s^{-1}$ (Worrall et al. 2007). Thus, using the HST data, and the accurate flux densities in the mid- and far-infrared derived in this paper, we are able to extend the
 Spectral Energy Distribution (SED) of the nucleus of NGC315 from the radio to X-ray  as shown in Figure~\ref{f:fig4}. 
Also shown in Fig. 4, is the SED of host galaxy which is seen to be very well fitted by Black-Body radiation of 3950K,
 which is consistent with its optical spectrum (Cid Fernandes et al. 2005)  and 
 the infrared colors derived  in this paper.
 
 Integrating the SED we obtain a bolometric luminosity of  $L_{\rm bol} = 1.93 \times 10^{43}$ ergs s$^{-1}$ for the
 nucleus of NGC 315.  Denicolo et al. (2005) obtained the central stellar velocity
 dispersion of NGC 315 to be $\sigma=246 \pm 8 \ \kms$, which according to the $M_{\rm BH}-\sigma$ relation (Tremaine et al.
 2002), corresponds to a central SMBH mass of $M_{\rm BH} \approx 3.1 \times
 10^{8} M_\odot$. Combining the SMBH mass with the bolometric luminosity and Eddington ratio $L/L_{\rm Edd} = 4.97 \times10^{-4}$, typical of low-luminosity AGNs (Ho, 1999).
 
 For comparison, we plotted in Fig.4 the SED of Sgr~A*,  the supermassive black hole 
 at the Galactic center, which is well measured from radio to X-rays (Yuan, 
 Quataert, \& Narayan 2003). We find that the SED shape of NGC~315 is remarkably similar 
 to that  of Sgr~A*, which might suggest a similar accretion mode - 
 radiatively inefficient accretion disk - for both NGC~315 and Sgr A*. This is surprising because  there are several systematic differences between Sgr A* and NGC~315: (1) L/L$_{\rm Edd}$ of Sgr A* is only  $3 \times 10^{-9}$
 (Yuan, Quataert, \& Narayan 2003),  more than 5 orders of magnitude lower than that 
 of NGC 315;  (2) The X-ray spectrum of Sgr A* is much steeper than that of NGC~315, 
 their photon indexes being 2.7 and 0.57, respectively (Baganoff et al. 2003; Worrall et al. 2007). 
 (3) 
 We see in Fig. 4 that the SED of NGC~315 deviates significantly from that of Sgr A* in the
 optical band.  Since the HST data has very high spatial resolution, the good agreement in the IR band may be fortuitous since we cannot separate emission 
 from the central dusty disk and AGN with the present  IRAC data. The match may also be fortuitous in the radio band since NGC~315 
 has extended radio jets, which may contaminate the core emission. 
 
 In order to throw more light on this subject, we also 
 plot in Fig. 4 the SED of NGC 4261, which has a similar L/L$_{\rm Edd}$ (2.8 $\times 10^{-5}$) 
 to NGC~315. We see that NGC~4261 matches  the SED of NGC~315 quite well 
 in the radio and optical bands, but is more luminous in X-rays. We must conclude that 
 the apparent match of SEDs does not provide a unique physical interpretation about the physics of radiation from AGN's. 

 In this work, we presented  {\it Spitzer} IRAC and MIPS observations of the elliptical
galaxy NGC~315. After careful removal of the dominant  stellar light from the host galaxy, 
 the IRAC revealed  that all the non-stellar IR emission originates from a compact region in the center of NGC 315. 
 Although there is a non-negligible contribution from the central dusty disk to the IR emission of 
 NGC~315,  the ring-like pattern in the non-stellar 8.0 \micron\ image, the significant excess 
 of dust mass  inferred from IR data than that of the optical dusty disk, and the power-law
 distribution of the IRAC fluxes strongly suggest  
 that we have  direct evidence of a 
 low-luminosity AGN in NGC 315, which might dominate the IR emission.
 
% Pozzi et al. (2007) selected 8
 %extremely red objects (EROs) in the redshift range of 0.9 $< z < $
 %2.08 with high X-ray flux density ($f_{2-10keV}> 10^{-14} erg
 %s^{-1} cm^{-2}$), high X-ray-to-optical ratio ($>$10), and extremely
 %red color ($R-K \ge 5$). 6 out of 8 objects show a typical de
% Vaucouleurs law for the K-band surface brightness distribution, and
% all of them are detected in the MIPS 24 \micron\ band.
% Neither the sensitivity nor spatial resolution of {\it Spitzer} IRAC is high enough to
% resolve the red nucleus in these elliptical galaxies at higher redshifts,  
% Pozzi et al. (2007) can
% not directly measure the SED of the active nucleus, instead they
% derived the SED by using a combination of an elliptical galaxy  template and torus emission, 
% and obtained
% the Eddington ratio of central black hole to be about  0.01 $\sim$ 0.03 for their sample,
% suggesting a low-accretion phase.
% We suggest that these mysterious ``E+24'' population found at $z \geq 1$ may be higher redshift analogs of NGC 315.

\acknowledgements
 The authors are very grateful to the anonymous referee for her/his 
 constructive report which improved the paper very
 much,  we also thank Jorge Melnick, Matthew Ashby and Luis Ho for 
 thoughtful discussion, and Feng Yuan
 for providing us the SED of Sgr~A*. This work has been 
 supported by Program for New Century Excellent Talents in University (NCET) of China.
 QGU would like to acknowledge the financial support from the China Scholarship
 Council (CSC),  the National Natural Science Foundation of China under
 grants 10221001 and 10633040, and  the National Basic Research Program 
 (973 program No. 2007CB815405).
 This research has made use of NASA's Astrophysics Data System Bibliographic
 Services and the NASA/IPAC Extragalactic Database (NED) which is operated
 by the Jet Propulsion Laboratory, California Institute of Technology, under
 contract with the National Aeronautics and Space Administration.
 This work is based on observations made with the {\it Spitzer Space Telescope},
 which is operated by the Jet Propulsion Laboratory, California Institute
 of Technology, under NASA contract 1407.

\clearpage

\clearpage

\clearpage

\clearpage
\begin{deluxetable}{lcccc}
  \tablewidth{0pt}
  \tablecaption{IRAC fluxes for NGC 315 \label{t:flux}}
  \tablehead{ \colhead{ } & \colhead{3.6$\mu$m} & \colhead{4.5$\mu$m} &  \colhead{5.8$\mu$m} & \colhead{8.0$\mu$m$^a$} }
 \startdata
 total   & $202.8 \pm 2.78$ & $120.8 \pm 2.12$ &  $73.9 \pm 1.83$ & $52.3 \pm 1.56$ \\
  nucleus & 2.424$\pm$0.291  & 3.650$\pm$0.357 & 5.870$\pm$0.453   & 11.450$\pm$0.633\\
 \enddata

\tablenotetext{a}{Fluxes are in units of mJy. }
\end{deluxetable}

\clearpage
\begin{deluxetable}{lcc}
  \tablewidth{0pt}
  \tablecaption{Data for the core of NGC 315 \label{t:sed}}
  \tablehead{ \colhead{$\nu$} & \colhead{$\nu F_{\nu}$} &   \\
           \colhead{(Hz)} & \colhead{(ergs cm$^{-2}$ s$^{-1}$) } & \colhead{Reference$^a$} }
\startdata
1.40 $\times$ 10$^9$   &  5.54 $\times$ 10$^{-15}$  & 1 \\
1.66 $\times$ 10$^9$   &  7.23 $\times$ 10$^{-15}$  & 2 \\
4.99 $\times$ 10$^9$   &  2.93 $\times$ 10$^{-14}$  & 2 \\
8.42 $\times$ 10$^9$   &  4.95 $\times$ 10$^{-14}$  & 2 \\
1.25 $\times$ 10$^{13}$ &  1.39 $\times$ 10$^{-11}$  & 3 \\
2.97 $\times$ 10$^{13}$ &  2.82 $\times$ 10$^{-12}$  & 4 \\
3.75 $\times$ 10$^{13}$ &  2.84 $\times$ 10$^{-12}$  & 3 \\
5.17 $\times$ 10$^{13}$ &  1.49 $\times$ 10$^{-12}$  & 3 \\
6.67 $\times$ 10$^{13}$ &  8.95 $\times$ 10$^{-13}$  & 3 \\
8.33 $\times$ 10$^{13}$ &  5.71 $\times$ 10$^{-13}$  & 3 \\
5.40 $\times$ 10$^{14}$ &  1.83 $\times$ 10$^{-13}$  & 5 \\
3.68 $\times$ 10$^{14}$ &  3.86 $\times$ 10$^{-13}$  & 5 \\
2.42 $\times$ 10$^{17}$ &  2.902 $\times$ 10$^{-13}$  & 6 \\
\enddata

\tablenotetext{a}{References: (1) Capetti et al. 2005. (2) Venturi et al. 1993.
(3) this work. (4) Heckman et al. 1983. (5) Verdoes Kleijn et al. 2002.
(6) Worrall et al. 2007. }
\end{deluxetable}
\end{document}